\documentclass[11pt]{article}
\usepackage{amsmath,amssymb,color,graphics,graphicx,amscd,amsfonts,epsf,ulem}
\usepackage{dsfont}
\newcommand{\one}{\ensuremath{\mathds{1}}}
\allowdisplaybreaks[4]

\date{\today}

\begin{document}

\begin{flushright} 

LLNL-JRNL-681857

UCB-PTH-16/01

SU-ITP-16/02

YITP-16-5
\end{flushright} 

\vspace{0.1cm}

\begin{center}
  {\LARGE

Chaos in Matrix Models and 
Black Hole Evaporation 

}
\end{center}
\vspace{0.1cm}
\vspace{0.1cm}
\begin{center}
	
Evan B{\sc erkowitz},$^a$\footnote{
	E-mail: berkowitz2@llnl.gov}
          Masanori H{\sc anada}$^{bcd}$\footnote
          {E-mail: hanada@yukawa.kyoto-u.ac.jp}
          and 
          Jonathan M{\sc altz}$^{eb}$\footnote
          {E-mail: jdmaltz@berkeley.edu or jdmaltz@alumni.stanford.edu}

\vspace{0.5cm}

$^a$ {\it Nuclear and Chemical Sciences Division, Lawrence Livermore National Laboratory, Livermore CA 94550, USA}

$^b${\it Stanford Institute for Theoretical Physics,
Stanford University, Stanford, CA 94305, USA}

$^c${\it Yukawa Institute for Theoretical Physics, Kyoto University,\\
Kitashirakawa Oiwakecho, Sakyo-ku, Kyoto 606-8502, Japan}	

$^d${\it The Hakubi Center for Advanced Research, Kyoto University,\\
Yoshida Ushinomiyacho, Sakyo-ku, Kyoto 606-8501, Japan}

 $^e$
{\it Berkeley Center for Theoretical Physics, University of California at Berkeley,\\
 Berkeley, CA 94720, USA}

\end{center}

\vspace{1.5cm}
%
%
\begin{abstract}
Is the evaporation of a black hole described by a unitary theory? 
In order to shed light on this question ---especially aspects of this question such as a black hole's negative specific heat---we consider the real-time dynamics of
a solitonic object in matrix quantum mechanics, which can be interpreted as a black hole (black zero-brane) via holography. 
We point out that the chaotic nature of the system combined with the flat directions of its potential naturally leads to
the emission of D0-branes from the black brane, which is suppressed in the large $N$ limit. 
Simple arguments show that the black zero-brane, like the Schwarzschild black hole, has negative specific heat, 
in the sense that the temperature goes up when it evaporates by emitting D0-branes. 
While the largest Lyapunov exponent grows during the evaporation, the Kolmogorov-Sinai entropy decreases. 
These are consequences of the generic properties of matrix models and gauge theory.  
Based on these results, we give a possible geometric interpretation of the eigenvalue distribution of matrices in terms of gravity. 

Applying the same argument in the M-theory parameter region, 
we provide a scenario to derive the Hawking radiation of massless particles from the Schwarzschild black hole. 
Finally, we suggest that by adding a fraction of the quantum effects to the classical theory, we can obtain a matrix model 
whose classical time evolution mimics the entire life of the black brane, from its formation to the evaporation.
\end{abstract}
\newpage
\section{Introduction}
\hspace{0.51cm}
Whether one can describe the formation and evaporation of a black hole from a unitary theory is the key question to answer in order to resolve 
Hawking's information paradox \cite{Hawking:1976ra}. This puzzle is attracting renewed interest after the proposal of the firewall paradox \cite{Almheiri:2012rt,Braunstein:2009my}. 
Gauge/gravity duality \cite{Maldacena:1997re,Itzhaki:1998dd} provides us with a stage on which to build a concrete setup: if the conjecture is correct, then we can translate the problem of black hole formation and evaporation into the real-time evolution of the dual gauge theory. 
In this paper, we consider the matrix model of M-theory, known as the BFSS matrix model \cite{Banks:1996vh,deWit:1988ig}.
Various objects, including a black hole, are regarded as bound states of D0-branes and open strings. 
A black hole is described by states where all the D0-branes and strings form a single bound state, described by generic non-commuting matrices and that all eigenvalues are clumped in the neighborhood of a point in space. 
In the 't Hooft large $N$ limit, where the matrix size $N$ is sent to infinity while the 't hooft coupling $g_{YM}^2 N$ is held fixed, at strong 't Hooft coupling, these states are conjectured \cite{Itzhaki:1998dd} to be dual to a black zero-brane. 
The deviation from strong coupling and large $N$ is conjectured to describe stringy corrections to the black hole. 
Previous thermodynamic simulations strongly suggested that the duality is valid not only at the supergravity level (large $N$ and strong coupling \cite{Anagnostopoulos:2007fw,Catterall:2008yz,Kadoh:2015mka,Filev:2015hia,Kabat:1999hp,Kabat:2000zv}) 
but also at the stringy level (finite-coupling \cite{Hanada:2008ez} and finite-$N$ \cite{Hanada:2013}). 
We will henceforth call this single bound state a ``black hole configuration" or a ``black zero-brane configuration".

This model has flat directions in its potential\footnote{Flat directions corresponds to configurations where $\sum_{M,M'}{\rm Tr}{[X_M,X_{M'}]^2} = 0$.} both classically and quantum mechanically \cite{de Wit:1988ct}. 
If we consider black hole configurations, 
this instability due to the flat directions is suppressed at large $N$ \cite{Anagnostopoulos:2007fw}. This is consistent with the dual gravity expectation that 
the black zero-brane becomes stable when the string coupling $g_s$ is set to zero. The instability can be understood as the Hawking radiation of 
D0-branes. Following the philosophy of the M-theory interpretation of this model \cite{Banks:1996vh}, the instability should translate to the emission of massless particles 
in the M-theory limit. 
(A similar process in the M-theory region has been studied in \cite{Banks:1997hz,Banks:1997tn,Banks:1997cm}, 
in order to explain the Schwarzschild black hole in eleven-dimensional supergravity.)

Most previous studies considered the BFSS model in imaginary time (Euclidean signature). 
In this paper, we interpret this phenomenon in real time (Minkowski signature). 
With currently available techniques, a full quantum real-time calculation of the matrix model is very difficult even numerically. 
However, since we are interested in the quantum nature of the gravity side 
we can exploit gauge/gravity duality to make use of the classical limit on the gauge theory side. The classical limit on the gauge theory side corresponds to a large $\alpha'$ correction on the gravity side as well as
the $1/N$ correction which exists in the gauge theory's classical limit. This correction controls the quantum correction on the gravity side. 
Thus, the classical dynamics of the matrix model can already tell us important lessons about the quantum gravity dynamics, 
although our argument is not restricted to the classical regime of the matrix model. 

In this paper, we first give a simple argument that the qualitative nature of the emission of a D0-brane from the black hole 
can naturally be described by the classical limit\footnote{Strictly speaking, a part of the quantum effects must be 
taken into account in order for the D0-brane to escape to infinity. This point will be made clear in the following sections.}; this is 
due to the existence of the flat direction in the matrix model and the chaotic nature of the system. 
When a D0-brane is emitted, {\it the number of degrees of freedom in the system decreases dynamically}, 
because open strings (off-diagonal elements of the matrices) decouple. 
Because of this, as D0-branes are emitted the temperature of the black hole goes up. 
We then show that the same argument holds for the strong coupling region where type IIA supergravity description is valid. 
We speculate that essentially the same mechanism also works in the M-theory region. 

This paper is organized as follows.
In Sec.~\ref{sec:matrix_model_and_BH}, we briefly review known results on 
black holes in the Matrix Model of M-theory. 
In Sec.~\ref{sec:BH_formation}, we review the formation of a black hole in the matrix model, 
putting an emphasis on a generic property of the matrix model important for understanding the evaporation process. 
Sec.~\ref{sec:evaporation} is the main part of this paper, where the evaporation is studied. 
Up to here, we restrict ourselves to the parameter region near the 't Hooft large $N$ limit. 
In Sec.~\ref{sec:M-theory}, we argue that the same argument in the M-theory parameter region might explain 
the properties of the Schwarzschild black hole. 
In Sec.~\ref{sec:effective_model}, we give a phenomenological model of the black zero-brane, 
whose classical dynamics can capture the qualitative features of the formation and evaporation of the black zero-brane. 
Finally in Sec.~\ref{sec:conclusion} we discuss our conclusions and refer to future work.

\section{The Matrix model and black holes}\label{sec:matrix_model_and_BH}
\hspace{0.51cm}
We consider maximally supersymmetric matrix quantum mechanics, 
which is the dimensional reduction of 4d ${\cal N}=4$ supersymmetric Yang-Mills theory to $0 + 1$ dimensions. 
Historically, this system was proposed as a non-perturbative 
formulation of a supermembrane (M2-brane) in eleven dimensional spacetime \cite{deWit:1988ig}. The connection to string/M-theory was first thought of as the low-energy effective action of D0-branes and open strings \cite{Witten:1995im} 
but was later understood as a formulation of M-theory in the infinite-momentum frame \cite{Banks:1996vh}.  
In Ref.~\cite{Itzhaki:1998dd}, this model was interpreted as the dual of a black hole (a black zero-brane in type IIA string theory 
or a Schwarzschild black hole in M-theory, depending on the parameter region) 
in the standard gauge/gravity dictionary. Clear reviews of matrix theory  can be found in \cite{Taylor:1999qk,Taylor:2001vb,Banks:1996my,Seiberg:1997ad,
Bigatti:1997jy,Banks:1999az}.

The Lagrangian is given by 
\begin{eqnarray}
L
=
\frac{1}{2g_{YM}^2}{\rm Tr}\Bigg\{
(D_t X_M)^2 
+
[X_M,X_{M'}]^2 
+
i\bar{\psi}^\alpha D_t\psi^\beta
+
\bar{\psi}^\alpha\gamma^M_{\alpha\beta}[X_M,\psi^\beta] 
\Bigg\},  
\end{eqnarray}
where $X_M$ $(M=1,2,\cdots,9)$ are $N\times N$ Hermitian matrices and $(D_tX_M)$ is the covariant derivative given by 
$(D_tX_M)=\partial_t X_M-i[A_t,X_M]$ and $A_t$ is the $U(N)$ gauge field. 
$\gamma^M_{\alpha\beta}$ $(M=1,2,\cdots,9)$ are $16\times 16$ matrices, which are the left-handed part of the gamma matrices in (9+1)-dimensions. 
$\psi_\alpha$ $(\alpha=1,2,\cdots,16)$ are $N\times N$ real fermionic matrices. 

This model has flat directions (directions where $\sum_{M,M'}{\rm Tr}{[X_M,X_{M'}]^2} = 0$) both classically and quantum mechanically. These flat directions have been identified with an instability of the supermembrane \cite{de Wit:1988ct}. 
Although this ``instability" is problematic when one tries to interpret the model as the first quantization of the membrane \cite{deWit:1988ig,de Wit:1988ct}, 
it makes the interpretation as the second-quantized M-theory possible \cite{Banks:1996vh}, because various block-diagonal matrix configurations, 
which are allowed due to the flat direction, can be interpreted as multi-object states \cite{Banks:1996vh}. 

In this paper, we follow the interpretation of this model proposed in Ref.~\cite{Itzhaki:1998dd}.  
There, the trivial vacuum (i.e. all eigenvalues of $X_M$ clump up near the origin) is considered 
from the viewpoint of gauge/gravity duality. 
In thermodynamics, the weakly-coupled type IIA superstring picture is valid at $N^{-10/21}\ll \lambda^{-1/3}T\ll 1$, 
while the M-theory picture becomes valid at much lower temperatures.  
The validity of the duality in the type IIA parameter region has been confirmed by numerical study of the matrix model 
\cite{Anagnostopoulos:2007fw,Hanada:2008ez,Hanada:2013,Catterall:2008yz,Kadoh:2015mka,Filev:2015hia}. 
(Reviews including the numerical methods can be found in \cite{Joseph:2015xwa,Hanada:2012eg,Catterall:2009it}.)
In the high-temperature region $\lambda^{-1/3}T\gg 1$, a weakly coupled dual gravity description does not exist.
Yet, since this region is expected to be smoothly connected to the low-temperature region without a phase transition 
(see Refs.~\cite{Anagnostopoulos:2007fw,Hanada:2008ez,Hanada:2013} for numerical results which support the absence of the transition), it can capture qualitative features of the black zero-brane solution of weakly coupled gravity.
For this reason, in the following, we call the trivial vacuum (the bunch of eigenvalues) a ``black hole configuration" or a ``black zero-brane configuration";
even when a weakly-coupled dual gravity description does not exist. 
We also refer to eigenvalues as ``D0-branes" when there is no risk of confusion.

If we consider black hole configurations,  
the instability due to the flat directions is suppressed at large $N$ \cite{Anagnostopoulos:2007fw}. This is consistent with the dual gravity expectation that 
the black zero-brane becomes stable when the string coupling $g_s$ is set to zero. The instability can be understood as Hawking radiation of 
D0-branes. 
Because the D0-brane is very heavy (its mass is of order $N$), it seems to have nothing to do with the usual Hawking radiation of massless particles at first sight. 
However, following the philosophy of the M-theory interpretation of this model -- `everything is made from D0-branes'-- \cite{Banks:1996vh}, the emission of D0-branes via the instability should translate to the emission of massless particles in the M-theory limit. 
Is it possible to see any resemblance of these features in type IIA and M-theory regions? 
This is the theme of this paper. 
\section{Black hole formation in the matrix model}\label{sec:BH_formation}
\hspace{0.51cm}
In this section we review the previously known results on the formation of a black hole in the matrix model, 
putting an emphasis on a generic property of the matrix model which turns out to be important in order to understand the evaporation process. 

Let us start with looking only at the classical dynamics of the model, which is justified at high temperature.\footnote{
The classical description is valid when open strings are much lighter than the temperature.
In black hole configurations the typical length of open strings scale as $(\lambda T)^{1/4}$ (see e.g. Ref.~\cite{Kawahara:2007ib}). 
Therefore the mass of the string, which is proportional to its length, becomes parametrically smaller than the temperature; hence the classical description is valid. 
When a D0-brane is emitted, however, open strings between the emitted D0 and the black hole becomes long and heavy; hence the classical approximation breaks down. This point will be revisited in Sec.~\ref{sec:why_evaporate}. 
}
The dynamics of fermions are negligible in this regime. 
So, we can consider only the bosonic part of the Lagrangian of the matrix model, which is given by 
\begin{eqnarray}
L
=
\frac{1}{2g_{YM}^2}{\rm Tr}\left(
\sum_{M}(D_tX_M)^2
+
\frac{1}{2}
\sum_{M\neq M'}
[X_M,X_{M'}]^2
\right). 
\end{eqnarray}

Near the 't Hooft large $N$ limit, this matrix model describes a system of $N$ D0-branes \cite{Witten:1995im}. 
When the $X_i$ are close to diagonal, the diagonal components $(X_1^{aa},\cdots,X_9^{aa})$ are regarded as the coordinates of 
$a$-th D0-brane in $\mathbb{R}^9$ and off-diagonal components represent the excitation of open strings connecting the D0-branes.  
A bound state of D0-branes, whose dual gravity description is a black hole (black 0-brane), 
is described by a set of highly non-commutative matrices. 
If the matrices are close to block-diagonal, it describes a multi-black hole state; each block describing a black hole.  

In the $A_t=0$ gauge, the classical dynamics are described by the equations of motion 
\begin{eqnarray}
\frac{d^2X_M}{dt^2}
-
\sum_{M'}[X_{M'},[X_M,X_{M'}]]
=
0
\label{eq:EOM}
\end{eqnarray}
along with the constraint 
\begin{eqnarray}
\sum_{M}\left[X_M,\frac{dX_M}{dt}\right]
=
0. 
\label{eq:constraint}
\end{eqnarray}

The classical time evolution of the matrix model is chaotic \cite{Savvidy:1982jk}. 
If we introduce a proper cutoff for matrix eigenvalues (i.e. putting the D0-branes in a box) so that the phase space becomes bounded, 
ergodicity should hold for generic initial conditions. Therefore, with the exception of very special initial conditions, almost all of phase space 
should be covered as the system evolves with time. 

What are the dominant (generic) configurations in the micro-canonical ensemble? 
A single black hole is entropically dominant; this can be seen from the following argument. The entropy of a sparse gas of $N$ D0-branes, 
which is described by a set of almost commuting matrices, scales only as $O(N)$
because it can be characterized by the positions of the D0-branes (diagonal elements of the matrices). Open strings (off-diagonal components) 
are long, heavy, and hence negligible compared to the diagonal components of the matrices. On the other hand, a black hole, which is a single bunch of the D0-branes (which is described by matrices where off-diagonal elements are not negligible), has an entropy of order $O(N^2)$. This is because open strings (off-diagonal elements) are excited. Multi-black hole states\footnote{Configurations with block diagonal matrices.} can also be considered. 
For example, two-black hole states $X_{\rm 2BH}$ have a block-diagonal structure
\begin{eqnarray}
X_{\rm 2BH}
=
\left(
\begin{array}{cc}
X' & 0 \\
0 & X''
\end{array}
\right), \label{eq:2BH}
\end{eqnarray}
where $X'$ and $X''$ are $N_1\times N_1$ and $N_2\times N_2$ are fully noncommutative matrices, with $N_1+N_2=N$. 
Because the number of degrees of freedom increases when off-diagonal blocks are excited, such block-diagonal configurations are less favorable compared to the single-block configurations.  
The spherical symmetry of the eigenvalue distribution can also be understood by an entropy argument.\footnote{In the case of nonzero angular momentum, this symmetry is reduced to a rotational symmetry.} 
 
The story is slightly different in our setup, in which the cutoff is not introduced. Because the phase space is not bounded, 
ergodicity does not make sense in the strictest sense. For example, if we consider D0-branes without any open string excitation
(i.e. all matrices are diagonal), there is no interaction and D0-branes just travel freely. 
Still, for generic configurations, all the D0-branes merge into a single bunch, and then, 
if the entropy of the black hole is large enough (i.e. $N$ is large enough), it is unlikely to see a large deviation from the black hole 
within a finite time, unless the initial configuration is fine-tuned. 
This has been confirmed in previous numerical studies \cite{Asplund:2011qj,Asplund:2012tg,Aoki:2015uha} 
and an analytic estimate has been given in Ref.~\cite{Gur-Ari:2015rcq}. 
We comment on this behavior further in Appendix~\ref{sec:D0_distribution_classical}.

To summarize: a black hole is formed because {\it the number of degrees of freedom increases dynamically when D0-branes 
clump up to form one bunch}\footnote{This can be thought of as un-Higgsing since the gauge symmetry is enhanced from $U(N_1)\times U(N_2) \rightarrow U(N_1 +N_2)$.}, and hence it is entropically favored. 
The same mechanism works at low-temperature region, both in the type IIA and M-theory regions. 
When the eigenvalues (D0-branes) come closer, off-diagonal elements (open strings) become lighter 
and the number of degrees of freedom increases dynamically; such configurations are favored entropically.
Previous imaginary-time simulations in type IIA region, which studied the canonical ensemble,
confirmed that the single-bunch configuration becomes more stable at larger $N$. 

In Sec.~\ref{sec:evaporation}, we show that the dynamical change of the number of degrees of freedom explains key features of 
black hole evaporation very naturally.  

\section{Black hole evaporation in the matrix model}\label{sec:evaporation}
\hspace{0.51cm}

\subsection{Chaos $+$  flat directions $\to$ evaporation}\label{sec:why_evaporate}
\hspace{0.51cm}
Because the classical matrix model is a chaotic theory with flat directions, even if a black hole is formed, 
it is possible that one of the D0-branes runs to infinity. This is Hawking radiation. 
Interestingly, although the emission rate is entropically suppressed, the emission after a long time is entropically favored. 
This rather counter-intuitive statement is explained as follows:
The entropy of the black hole (BH) is $N^2$. Separating a D0 brane from the black hole leaves the entropy of BH$+$D0 to scale as $(N-1)^2$, 
because open strings stretched between BH  and D0 decouple dynamically. 
The suppression factor is then $\sim e^{-N}$. Hence, evaporation is suppressed at large $N$.
However, once D0-brane goes sufficiently far from the BH, then it can move freely, 
which adds $\log V$, where $V$ is the volume of the space, to the entropy. 
Unless we introduce a cutoff artificially, $V$ is infinite. Therefore, the radiation comes out after long time. 
Note that the time scale for the evaporation, $e^{+N}$, is smaller than the recurrence time $e^{+N^2}$ of the black hole and 
hence it is physically meaningful. 

In fact this scenario must be made a little bit more precise, because the flat direction is very narrow in the classical matrix model.  
Indeed, as we show in Appendix \ref{sec:D0_distribution_classical}, the distribution of the largest eigenvalue $r$ of 
$R\equiv\sqrt{\sum_M X_M^2}$, which can be regarded as the radial coordinate of the emitted D0-brane,
 behaves as $\rho(r)\sim (\lambda T)^{-1/4}\cdot [r/(\lambda T)^{1/4}]^{-8(2N-3)}$.  
Therefore, no mater how large of a value it rolls to it always comes back, unless the initial condition is infinitely fine-tuned.  
(This suppression is the quantitative version of the suppression factor $\sim e^{-N}$ mentioned above). 
This behavior is tied to the breakdown of the classical treatment of the matrix model at large $r$. 
The classical treatment is only valid at sufficiently high energy, or equivalently, at high temperature. 
Here `high temperature' means that the temperature $T$ times the Boltzmann constant is higher than the energy quanta, 
typically characterized by the mass of the open strings between D0-branes, so that a lot of open strings are excited.  
When the eigenvalues are separated, however, the open strings become heavier and this condition breaks down, so that quantum effects become important. 
When the distance is long enough, the one-loop calculation becomes valid and then one can show that 
the attractive and repulsive forces between separated D0-branes cancel. The cancellation comes from bosonic and fermionic degrees of freedom of the matrix model; this includes the zero-point energy contributions. This cancellation is not captured by the purely classical model which explains the return of the eigenvalue.
Hence, it would make sense to adopt a crude approximation: 
once one of the D0-branes travels far enough away from the others the interaction is turned off. 
This crude approximation is already good enough to realize the evaporation. 

The same entropy argument applies to type IIA region as well, with a similar emission rate $\sim e^{-N}$, 
because the $N$-dependence is the same. (The dependence on the temperature and coupling can change.)

\subsection{A black hole heats up as it evaporates}\label{sec:specific_heat}
\hspace{0.51cm}

In this section we show that the black hole described in the matrix model has negative specific heat, 
in the sense that the temperature of the black hole goes up when a D0-brane is emitted.  
For this purpose, we consider the matrix configurations corresponding to one black hole $X_{\rm BH}$ and 
`black hole + D0' $X_{\rm BH+D0}$. The former is fully noncommutative, 
while the latter has a block-diagonal structure  
\begin{eqnarray}
X_{\rm BH+D0}
=
\left(
\begin{array}{cc}
X' & 0 \\
0 & x_{\rm D0}
\end{array}
\right), 
\end{eqnarray}
where $X'$ is an $(N-1)\times (N-1)$ fully noncommutative matrix. 
This is a special case of the `two-black hole configuration' in \eqref{eq:2BH}. 
We can consider various processes: a merger of two black holes to a single black hole, $X_{\rm 2BH}\to X_{\rm BH}$;  
growth of a black hole by absorption of a D0-brane, $X_{\rm BH+D0}\to X_{\rm BH}$; 
and the time-reversed process, emission of a D0-brane from a black hole, $X_{\rm BH}\to X_{\rm BH+D0}$. This process of emission can be thought of as Higgsing the gauge symmetry, as the gauge group goes from $U(N)\rightarrow U(N - 1)\times U(1)$.
In the matrix model, the energy of the entire system is conserved, 
and hence\footnote{Here we assume eigenvalues of $X'$ and $x_{\rm D0}$ are sufficiently separated in ${\mathbb R}^9$, 
so that the contribution from the off-diagonal blocks is negligible including the quantum correction.} 
\begin{eqnarray}\label{energyrel}
E_{X_{\rm BH}}
=
E_{X'}+E_{\rm D0}. 
\end{eqnarray}
\subsubsection{High-temperature region}
\hspace{0.51cm}
Let us consider the high temperature region, where the classical approximation is valid. 
There, by equipartition, the energy of the noncommutative block is the temperature times the number of degrees of freedom. 
Hence energy conservation \eqref{energyrel} leads to \footnote{
Here we ignored the motion of the center of mass of the black hole. 
This treatment is parametrically good when $N$ is large, because it gives only a negligible correction to $E_{X'}$
when $E_{\rm D0}$ is of order $N^0$.  Note that we assume the center of mass of entire system is at rest. }
 
\begin{eqnarray}
cN^2T
=
c(N-1)^2T'+E_{\rm D0}, 
\end{eqnarray}
where $c=6$ is a numerical factor obtained by counting the number of degrees of freedom 
and using the virial theorem (see e.g. Eq.~6 in Ref.~\cite{Gur-Ari:2015rcq})\footnote{$c=6$ is the large N value. At finite $N$ there is a correction of $O(1/N^2)$  which gives $c = 6 -\frac{33}{N^2}$. This is further explained in Appendix \ref{sec:classical_dof}. This  correction does not change our argument in the large $N$ limit.}.
In the limit we are considering, $T$, $T'$ and $E_{\rm D0}$ are $O(N^0)$.\footnote{
Here we did not count the rest-mass energy, because the gauge theory describes the excitation above zero temperature. }$^{,}$\footnote{
Note that $E_{\rm D0}$ can become as large as $O(N)$ or $O(N^2)$ in principle, but such extreme configurations 
should be suppressed entropically, because the entropy of the black hole becomes smaller 
as the energy carried away by the emitted D0-brane becomes larger.
We thank Y.~Hyakutake for a discussion concerning this point. 
}
Then, 
\begin{eqnarray}
T
=
\left(1-\frac{1}{N}\right)^2T'+\frac{E_{\rm D0}}{cN^2}
=
T'-\frac{2}{N} T'+\frac{T'+c^{-1}E_{\rm D0}}{N^2},   
\end{eqnarray}
and hence
\begin{eqnarray}
T<T'\simeq\left(1+\frac{2}{N}\right)T. 
\end{eqnarray}
The details of the energy spectrum of the emitted D0 does not affect the temperature of the black hole;  
the energy per degree of freedom increases because the number of degrees of freedom decreases dynamically. 

\subsubsection{Low-temperature region}
\hspace{0.51cm}
Next let us consider the low-temperature region, where the dual type IIA calculation \cite{Itzhaki:1998dd} is justified. 
For a single-black hole configuration, the energy is given by\footnote{
Strictly speaking, this relation, which is derived for the canonical ensemble, is applicable to the current case only at large $N$. 
Here we assume $N$ is sufficiently large.  
} 
\begin{eqnarray}
E_{X_{\rm BH}}
=
c'\lambda^{-3/5}T^{14/5}N^2, 
\end{eqnarray}
where $c'\simeq 7.4$ is a numerical factor calculable by using the dual gravity prescription. 

When one D0-brane is emitted, the remaining black hole is described by $U(N-1)$ gauge group. 
Then the 't Hooft coupling should be modified to 
\begin{eqnarray}
\lambda'=g_{YM}^2(N-1) = \frac{N-1}{N}\lambda. 
\end{eqnarray}
Because D0-branes went infinitely far from the black hole by assumption, 
we can relate the energy and temperature of the black hole by using the formula for a single black hole
as\footnote{
Again, we ignored the motion of the center of mass of the black hole. 
This treatment is parametrically good when $N$ is large. 
} 
\begin{eqnarray}
E_{X'}
=
c'\lambda'^{-3/5}T'^{14/5}(N-1)^2. 
\end{eqnarray}
Therefore from the energy conservation (\ref{energyrel}) we obtain 
\begin{eqnarray}
T^{14/5}
=
T'^{14/5}\left(1-\frac{1}{N}\right)^{7/5}
+
\frac{\lambda^{3/5}E_{\rm D0}}{c'N^2}, 
\end{eqnarray}
and hence 
\begin{eqnarray}
T<T'\simeq \left(1+\frac{1}{2N}\right)T, 
\end{eqnarray} 
regardless of the exact value of the energy of the emitted D0-brane. 
Again in this case, the black hole heats up.
\subsection{Lyapunov exponent and Kolmogorov-Sinai entropy}
\hspace{0.51cm}
Because the time scale for the D0-brane emission $\sim e^{+N}$ is much larger than the scrambling time $\sim\log N$ 
(see e.g. Refs.~\cite{Gur-Ari:2015rcq,Sekino:2008he,Shenker:2013pqa,Shenker:2014cwa,Maldacena:2015waa}), 
the Lyapunov exponent can make sense despite the instability. 
The largest Lyapunov exponent of a black hole state is proportional to $T$ in the IIA-string region \cite{Shenker:2013pqa,Shenker:2014cwa,Maldacena:2015waa}, and hence it increases as the D0-brane is emitted and the temperature goes up.  
At high-temperature, the largest exponent scales as $(\lambda T)^{1/4}$ \cite{Gur-Ari:2015rcq}. 
When the D0-brane is emitted, the 't Hooft coupling effectively changes as $\lambda\to\frac{N-1}{N}\lambda$, 
while the change of the temperature is $\Delta T\simeq \frac{2T}{N}$, and hence the exponent increases to 
$\left(1+\frac{1}{4N}\right)(\lambda T)^{1/4}$.

The Kolmogorov-Sinai (KS) entropy, which is given by the sum of all positive Lyapunov exponents\footnote{
Precise mathematical definition of the KS entropy in quantum theory is a subtle issue. 
Here we assume that the equality which holds at classical region is valid in quantum region as well.}, 
is a better quantity to characterize the strength of the chaos. 
At high temperature, the KS entropy of the $N\times N$ matrix configurations is proportional to $N^2(\lambda T)^{1/4}$ \cite{Gur-Ari:2015rcq}.  
After the emission of a D0-brane, it changes to $(N-1)^2\left(1+\frac{1}{4N}\right)(\lambda T)^{1/4}\simeq \left(N^2-\frac{7N}{4}\right)(\lambda T)^{1/4}$. 
Hence the KS entropy decreases. The point here is that the growth of each Lyapunov exponent 
cannot overcome the decrease of the number of degrees of freedom. Note that perturbation of the off-diagonal elements 
does not lead to chaos, because off-diagonal elements behave as harmonic oscillators when the separation is large.   

At low temperature, although the full Lyapunov spectrum has not been calculated, it is reasonable to assume that the spectrum is degenerate 
in the supergravity limit ($T\to 0$), 
because otherwise a nontrivial correction should appear in the out-of-time-order correlation function \cite{Shenker:2013pqa,Shenker:2014cwa,Maldacena:2015waa}. 
Then the KS entropy is proportional to $N^2T$. 
After emitting a D0, it changes to $(N-1)^2(T+\Delta T)$, where $\Delta T\simeq \frac{T}{2N}$, 
and hence $(N-1)^2(T+\Delta T)\simeq \left(N^2-\frac{3N}{2}\right)T<N^2T$. 
Again, the KS entropy decreases. 

We can repeat essentially the same calculation for a merger of black holes, such as $X_{\rm 2BH}\to X_{BH}$.  
There, the KS entropy after the merger is larger than the sum of the KS entropies of two black holes before the merger. 
Therefore, if we use the KS entropy as a measure of the speed of the scrambling, a bigger black hole is a faster scrambler. 
 
\subsection{A possible geometric interpretation of the eigenvalue distribution}
\begin{figure}[htbp]
\begin{center}
\rotatebox{0}{
\scalebox{0.4}{
\includegraphics{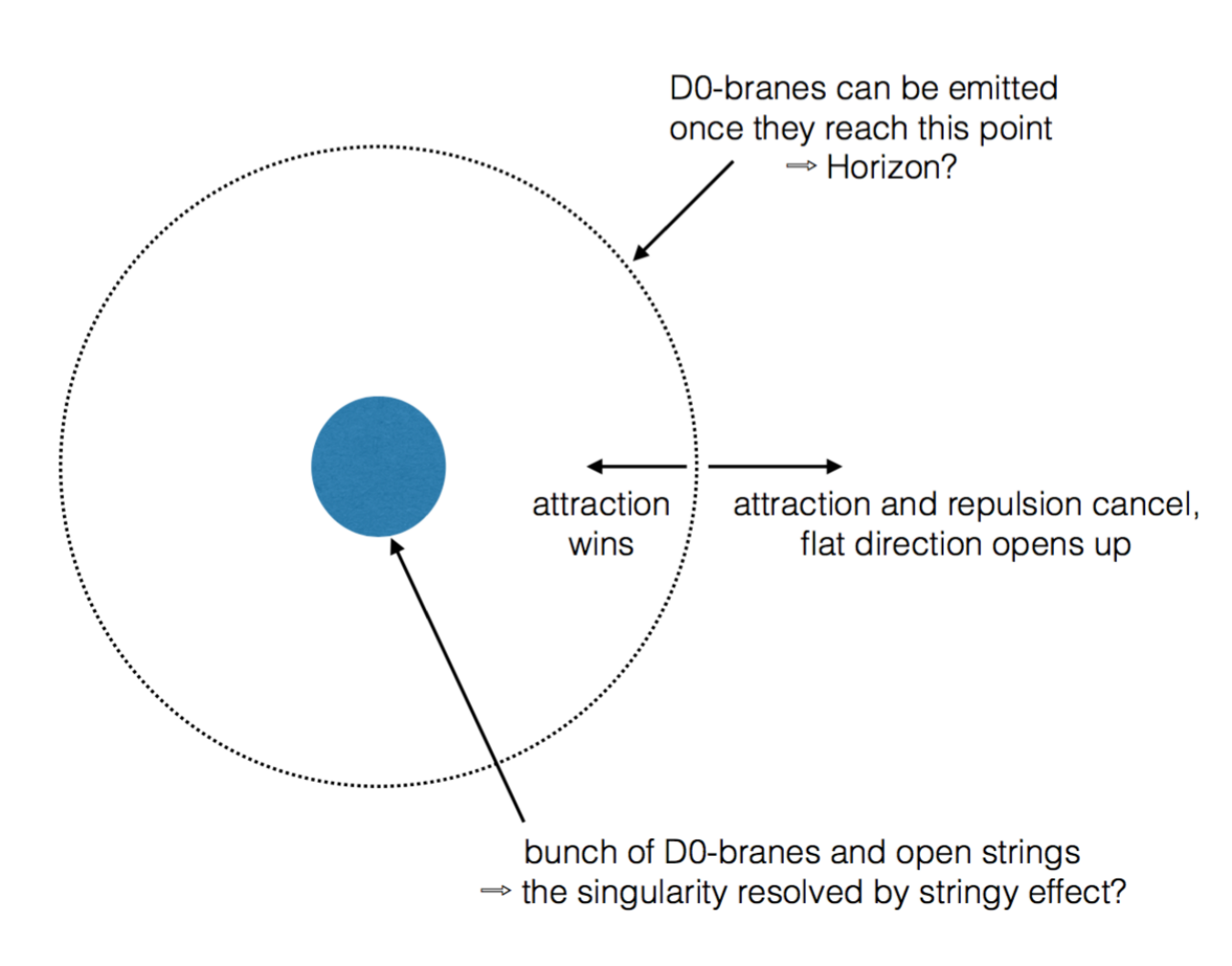}}}
\end{center}
\caption{A possible geometric interpretation of the eigenvalue distribution.
}\label{fig:geometric_interpretation}
\end{figure}

\hspace{0.51cm}
It is important to understand how the geometry of the gravitational theory is encoded in the matrices. 
In the previous sections, we showed that the D0-branes can be emitted from the black zero-brane, 
once they reach the point where the flat direction opens up. On the gravity side, it seems that particles escape at this distance. 
It reminds us of the tunneling picture of the Hawking radiation \cite{Parikh:1999mf}. 
Hence it would be reasonable to identify this point (where the flat direction opens up) with the horizon\footnote{This distance might be anywhere in the near horizon zone.
} (Fig.~\ref{fig:geometric_interpretation}). 
In this interpretation, the bunch of D0-branes and open strings are sitting behind the horizon, and there the notion of the `coordinate' of the D0-brane is obscure. 
It may then be possible to interpret the central bunch as the singularity resolved by the stringy effects.

\section{M-theory region and massless Hawking radiation}\label{sec:M-theory}
\hspace{0.51cm}
So far we have considered the parameter region near the 't Hooft large $N$ limit, where $g_{YM}^2\sim 1/N$, in the microcanonical ensemble. 
There, the mass of D0-branes are $1/g_s\sim 1/g_{YM}^2\sim N$. Due to this the emission of massless particles cannot be described
by the dynamics of eigenvalues, and 
the emission of D0-branes are suppressed exponentially, $e^{-1/g_{YM}^2}\sim e^{-N}$.  
At sufficiently large $N$, the single black hole configuration is rather stable and it makes sense to restrict the path integral there; 
or in other words, the canonical ensemble can reasonably be restricted to {\it single black hole}  states. 
Then the specific heat in the canonical ensemble, for {\it single black hole} configurations, is positive. 
This is consistent with the dual gravity prediction---the black zero-brane has a positive specific heat in the canonical ensemble, 
when the emission of D0-brane is ignored. 
If we measure the mass of the meta-stable bound state in the Monte Carlo simulation of the imaginary time theory, 
single black hole states should give the dominant contribution---consistent with previous simulation results
\cite{Anagnostopoulos:2007fw,Catterall:2008yz,Kadoh:2015mka,Filev:2015hia,Hanada:2008ez}. 

In this section, we consider the M-theory limit: $g_{YM}^2\gtrsim O(N^0)$ and $T$ fixed. 
The Schwarzschild black hole in M-theory behaves rather differently from the black zero-brane. It emits massless particles, the radiation rate is suppressed only by a power of the black hole mass, 
and the specific heat in the canonical ensemble is negative. 
Still, these properties might be described by the matrix model as follows.  

Firstly, because $g_{YM}^2\gtrsim O(N^0)$, the mass of a D0-brane $\sim 1/g_{YM}^2$ does not increase with $N$ any more, 
and hence the emission of massless particles 
can be described by the dynamics of eigenvalues. 
The suppression factor $e^{-1/g_{YM}^2}$ becomes of order one, 
which means the emission is not suppressed much. This is consistent with the fact that the radiation rate from the Schwarzschild BH 
is suppressed only by a negative power of the mass. 

Furthermore, because the instability is rather large, it would not make sense to distinguish 
{\it single black hole}  configurations with others very sharply --  
if an eigenvalue can easily fluctuate to large distance from the bunch of other eigenvalues, how can we demarcate the border between
`inside' and `outside' the bunch? 
So, when we try to pick {\it single black hole} configurations in the canonical ensemble (for example by putting it in a box or 
selecting meta-stable single-bunch configurations in the Monte Carlo simulation), 
other states (e.g. something between ``a single black hole" and ``a single black hole plus radiation") can mix into the ensemble. 
Those other states typically have lower energy than ``a single black hole", because they have fewer degrees of freedom,\footnote{
As an extreme example, the sparse gas of D0-branes has only $O(N)$ degrees of freedom, and hence the energy of the gas state should be suppressed 
when compared to a single-BH state by a factor of $N$. 
} 
and hence lower the energy expectation value of the whole ensemble. 

In a previous Monte-Carlo study of the imaginary time theory \cite{Hanada:2013}, it has been observed that the instability is severe 
at a $N$-dependent interval $(T_{N,L}, T_{N,H})$, 
where $T_{N,L}$ increases with $N$ and $T_{N,H}$ decreases with $N$.\footnote{
That smaller $N$ is less stable should be justified by the entropy argument. The stability at lower $T$ is related to the attraction 
coming from the diagonal element of fermionic zero modes \cite{Aoki:1998vn}. } 
This suggests that if we increase the temperature in the M-theory region (at very low temperature, below $T_{N,L}$) 
the instability increases. Then ``other" states can contribute more, and the energy would go down as temperature increases. 
It would be possible to explain the negative specific heat of the Schwarzschild black hole in the canonical ensemble in this manner. 

\section{A phenomenological model for a real-time numerical simulation}\label{sec:effective_model}
\begin{figure}[htbp]
\begin{center}
\rotatebox{0}{
\scalebox{0.6}{
\includegraphics{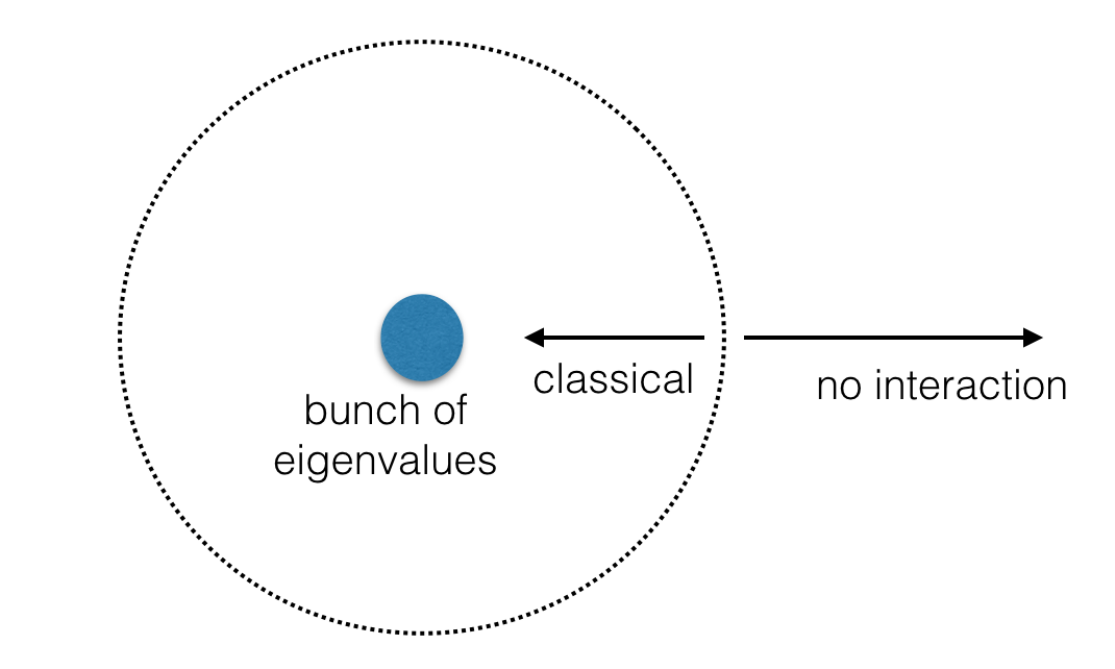}}}
\end{center}
\caption{A phenomenological model of the black zero-brane / Black Hole. 
}\label{fig:effective_model}
\end{figure}

\hspace{0.51cm}
As we have seen, the BFSS matrix model describes the evaporation of a black hole to some extent 
already at the level of the classical equations of motion. In order for the `evaporation' to take place, 
the chaos and the flat direction are the keys, and other details do not matter. 
In principle, the flat direction can be correctly treated by taking into account the quantum correction order by order. 
Here, as a phenomenological treatment, let us turn-off the interaction between an emitted D0-brane and the {\it black hole} (bunch of other D0-branes)
once the emitted D0 goes beyond a threshold (Fig.~\ref{fig:effective_model}). 

One way to achieve this is to 
take the basis in which $R\equiv \sqrt{X_M^2}$ is diagonal, $URU^{-1}={\rm diag}(r_1,r_2,\cdots,r_N)$, 
where $r_1\le r_2\le\cdots\le r_N$, and set to zero the off-diagonal elements of the $N$-th row and column of $UX_MU^{-1}$
when $r_N$ exceeds a cutoff value. 
This can be done in a covariant manner as follows. 
Let $f(x)$ be a function, which is close to 1 at $x\lesssim 1$ and goes to zero at $x\gtrsim 1$, 
for example $f(x)=e^{-x^{k}}$ with a large enough integer $k$.  
(Note that it becomes a step function when $k\rightarrow\infty$.)  Then, by replacing matrices as 
\begin{eqnarray}
X_M 
\quad\to\quad
\tilde{X}_M\equiv f(R/L)\cdot X_M\cdot f(R/L)
+
(\one_N-f(R/L))\cdot X_M\cdot (\one_N-f(R/L))
\end{eqnarray}
and 
\begin{align}
D_t X_M 
\quad\to\quad
\widetilde{D_tX_M}\equiv f(R/L)&\cdot (D_t X_M)\cdot f(R/L)
\\&+
(\one_N-f(R/L))\cdot (D_t X_M)\cdot (\one_N-f(R/L))
\nonumber\\
\end{align}
we can turn-off the off-diagonal elements when $r_N>L$. 
One may also add a very small mass term, 
\begin{eqnarray}
m^2{\rm Tr}\tilde{X}_M^2, 
\end{eqnarray}
in order to prevent the random walk of the black hole's center of mass. As long as $m^2L^2\ll 1$, this deformation does not prevent D0 emission. 
Note that this model is crude; there is an ambiguity for a choice of the cutoff procedure, including the choice of $f$, 
and the temperature dependence of cutoff value $L$ is ignored here. 
Although this model does not necessarily capture the quantitative features of the full theory, 
for example the details of the spectrum of the emitted particles,  
we can expect that it can mimic the life of a black hole to some extent: the formation and thermalization can take place 
at the classical region as in the previous studies \cite{Asplund:2011qj,Asplund:2012tg,Aoki:2015uha}, 
and then at the later time D0-branes are emitted one by one. 
We hope to report the simulation results on this phenomenological model, as well as results on more systematic approximations 
of the BFSS matrix model, in future publications.

\section{Conclusion and discussion}
\label{sec:conclusion}
\hspace{0.51cm}
In this paper we illustrated how the evaporation of a black hole can be described by the matrix model of M-theory. 
We considered a property in the 't Hooft large $N$ limit, which is dual to type IIA superstring theory, 
and pointed out that the combination of chaos and flat directions can lead to the evaporation of the black zero-brane. 
A generic property of the matrix models---the dynamical change of the number of degrees of freedom---causes the black zero-brane to become hotter as it evaporates by emitting D0-branes. 
We have also seen that the largest Lyapunov exponent increases during the evaporation, 
while the Kolmogorov-Sinai entropy decreases. 
Based on these results, we gave a few speculations in Sec.~\ref{sec:M-theory} and Sec.~\ref{sec:effective_model}. 

Rather surprisingly, the classical dynamics already contains all the essence of black hole evaporation, 
modulo the subtlety that the flat direction is not sufficiently wide. 
This problem can be cured by including a one-loop quantum correction. 
Therefore, it should be possible to gain useful insight into the Hawking radiation  
by starting with the classical theory and adding the quantum effects order by order in the coupling constant. 
The first and very important step would be the numerical confirmation of the thermal spectrum.

\section*{Acknowledgements} 
\hspace{0.51cm}

The authors would like to thank Sinya Aoki, David Berenstein, Ori Ganor, Guy Gur-Ari, Masahiro Hotta, Yoshifumi Hyakutake, So Matsuura, 
Paul Romatschke, Stephen Shenker, Hidehiko Shimada and Leonard Susskind for stimulating discussions and comments.

The work of M.~H. is supported in part by the Grant-in-Aid of the Japanese Ministry 
of Education, Sciences and Technology, Sports and Culture (MEXT) 
for Scientific Research (No. 25287046). 
The work of J.~M. is supported by the California Alliance fellowship (NSF grant 32540).

This work was performed under the auspices of the U.S. Department of Energy by Lawrence Livermore National Laboratory under Contract DE-AC52-07NA27344.  Numerical calculations were performed on the Aztec cluster at LLNL, supported by the LLNL Multiprogrammatic and
Institutional Computing program through a Tier 1 Grand Challenge award.
\vspace{0.25in}
\appendix
\begin{flushleft}
{\Huge \textbf{Appendix}}
\end{flushleft}
\section{Number of dynamical degrees of freedom in classical matrix model}\label{sec:classical_dof}
\hspace{0.51cm}
In this appendix we explain how to obtain the number of degrees of freedom in the classical theory explicitly. 

Consider a matrix model with $d$ matrices $X_1,X_2,\cdots,X_d$, each containing $N^2$ real degrees of freedom.
Firstly let us assume $N^2$ is greater than $d$.
We should subtract the number of constraints from these $d N^2$ matrix elements in order to determine the number of degrees of freedom. 
Taking into account the Gauss law constraint and the residual gauge symmetry, we should subtract $N^2-1$. 
Also we should take into account the conservation of momentum ${\rm Tr}V_M$ and angular momentum 
${\rm Tr}(X_M V_{M'}-X_{M'} V_{M})$, which further subtracts $d$ and $\frac{d(d-1)}{2}$ respectfully, from the number of elements. 
Then the number of degrees of freedom is $(d-1)\left(N^2-1-\frac{d}{2}\right)$. 

When $d\ge N^2$, the situation becomes a little bit more complicated. 
The model describes motion of $N$ D0-branes and $N(N-1)$ open strings in $d$-dimensional space, 
by using $d$ ``coordinates." Therefore, when $d\ge N^2$, there must arise constraints to preserve the fact the $d$ ``coordinates" are still describing a theory of $N^2-1$ matrices. 

\section{D0-brane distribution in classical matrix model}\label{sec:D0_distribution_classical}
\hspace{0.51cm}
We consider a classical $d$-matrix model described by the Lagrangian 
\begin{eqnarray}
L
=
\frac{1}{2g_{YM}^2}{\rm Tr}\left(
\sum_{i}(D_tX_i)^2
+
\frac{1}{2}
\sum_{i\neq j}[X_i,X_j]^2
\right),   
\end{eqnarray}
where $X_1,\cdots,X_d$ are $N\times N$ traceless Hermitian matrices. 
In Ref.~\cite{Gur-Ari:2015rcq}, a theoretical prediction on the distribution of the largest eigenvalue $r$ of $R\equiv\sqrt{\sum_M X_M^2}$
in the $A=0$ gauge, 
for generic values of $d$ and $N$, is given. Here we review the argument in Ref.~\cite{Gur-Ari:2015rcq} and 
confirm the prediction by numerical calculation. 

Suppose $R\simeq {\rm diag}(0,0,\cdots,r)$. 
Then, by using $SO(d)$, we can take 
\begin{eqnarray}
X_M
&=&
{\rm diag}(0,0,\cdots,0) + {\rm small\ fluctuation}\ (M=1,2,\cdots,d-1), 
\nonumber\\
X_d
&=&
{\rm diag}(0,0,\cdots,0,r) + {\rm small\ fluctuation}. 
\end{eqnarray}
When $r$ is sufficiently large, the potential energy coming from open strings (i.e. $N$-th row and column) 
can be approximated by $\frac{N}{\lambda}\sum_{M=1}^{d-1}\sum_{i=1}^{N-1}r^2|X^M_{Ni}|^2$. Because this must be smaller than the total energy, 
we have 
\begin{eqnarray}
\sum_{M=1}^{d-1}\sum_{i=1}^{N-1}|X^M_{Ni}|^2
<
\frac{\lambda E}{Nr^2}
\sim
\frac{\lambda NT}{r^2}.  
\label{loose_bound}
\end{eqnarray}
This is actually a very loose bound, though it is sufficient to show the absence of a flat direction. 
Equation \eqref{loose_bound} can actually be replaced by 
\begin{eqnarray}
\sum_{M=1}^{d-1}\sum_{i=1}^{N-1}|X^M_{Ni}|^2
<
\frac{(\lambda E/N)}{Nr^2}
\sim
\frac{\lambda T}{r^2},   
\label{tight_bound}
\end{eqnarray}
because typically $O(N)$ d.o.f. can carry only $O(N)$ energy. More precisely speaking, 
configurations with larger energy are entropically suppressed.

Because of this, phase space volume at $[r,r+dr]$ has a suppression factor 
\begin{eqnarray}
dr\cdot r^{d-1} 
\left(\frac{1}{r^2}\right)^{(d-1)(N-1)},   
\end{eqnarray}
where $r^{d-1}$ comes from the $SO(d)$ angular degrees of freedom. 
Except for $d=2, N=2$, the integral with respect to $r$ converges, and hence the large $r$ region is suppressed; 
the possibility that $r$ becomes larger than $R$ is suppressed by $R^{-(d-1)(2N-3)+1}$. 
The temperature dependence appears only through the combination $\lambda T$, 
and can be determined by dimensional analysis to be
\begin{eqnarray}\label{eq:dimensional_analysis}
\rho(r)
\sim
(\lambda T)^{-1/4}\cdot 
\left(\frac{r}{(\lambda T)^{1/4}}\right)^{-(d-1)(2N-3)}.  
\end{eqnarray}
Because of this suppression, the phase space has a finite volume except for $d=2,N=2$.\footnote{
A similar argument has been applied to the zero-dimensional matrix model, in order to show the finiteness of the partition function \cite{Austing:2001bd}.
}

Two remarks are in order here. Firstly, the remaining bunch of $N-1$ D0-branes considered above is more stable at larger $N$. 
When $N$ is too small, the bunch can become unstable and more D0-branes will escape from the bunch, and then the above estimate may change. 
We can expect the estimate becomes parametrically good at large $N$. ($N=2$ is the exception, because the remaining ``bunch" is already 
one D0-brane, which does not have further instability.) 
Secondly, when $d\ge N^2$, $X_M$ with $M\ge N^2$ are not independent, and hence the counting changes, 
as we have seen in Appendix~\ref{sec:classical_dof}. 

\begin{figure}[htbp]
\begin{center}
\includegraphics[width=0.75\textwidth]{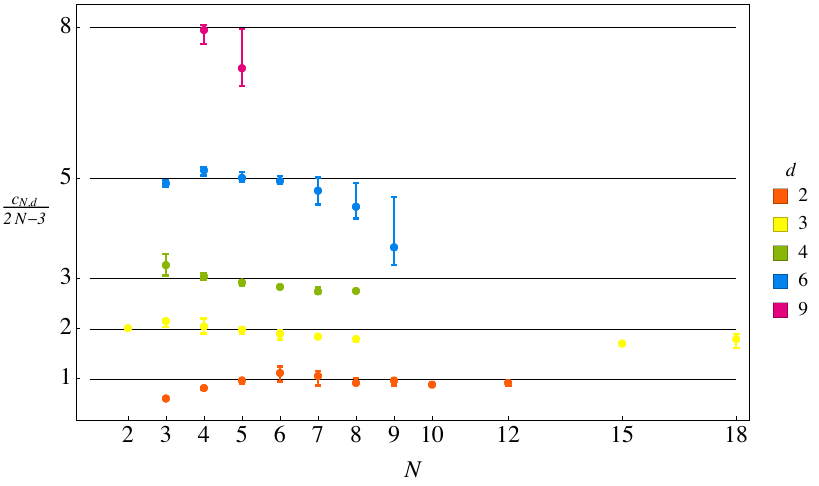}
\end{center}
\caption{
The observed numerical value of $c_{N,d}/(2N-3)$ as a function of $N$, for several values of $N$ and $d$.  Thin black lines indicate where $c_{N,d}=(d-1)(2N-3)$ as in \eqref{eq:dimensional_analysis}.  We fit a line to every span of points to the right of the maximum of the corresponding numerical distribution and constructed a weighted histogram according to goodness of fit.  The best values correspond to the maximum of that histogram, while the error bars were determined by finding where that histogram fell by a factor of $e$ to either side.
}\label{fig:exponent_vs_n}
\end{figure}

\begin{figure}[htbp]
\begin{center}
\includegraphics[width=0.75\textwidth]{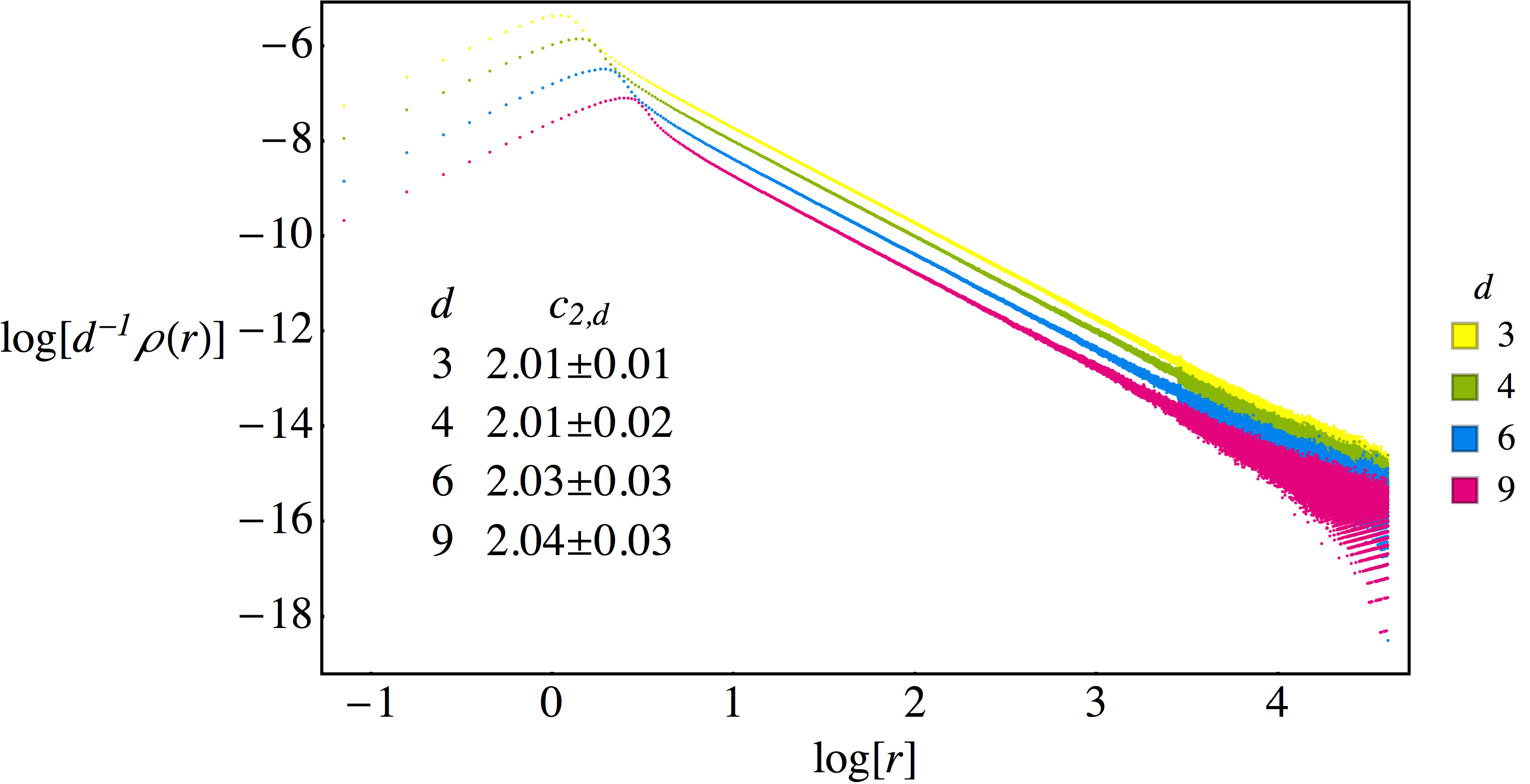}
\end{center}
\caption{
The observed numerical distribution as a function of $r$ for $N=2$, $d=3,4,6,9$. The powers $c_{2,d}$ take the same values within error.  We divide by $d$ simply to visually offset the different distributions from one another.
}\label{fig:histograms_vs_r_Nc2}
\end{figure}

In order to test this analytic estimate, we have performed a numerical analysis, 
using the discretization method employed in Refs.~\cite{Aoki:2015uha,Gur-Ari:2015rcq}. 
The tail of the distribution of $r$ can be fit by a power law $r^{-c_{N,d}}$. 
In Fig.~\ref{fig:exponent_vs_n}, the ratios $c_{N,d}/(2N-3)$ are plotted for various $N$ and $d$ which satisfy $d<N^2$. 
The values agree well with $d-1$, except for $N=3$. As we have explained above, a large deviation at $N=3$ 
is not surprising. In Fig.~\ref{fig:histograms_vs_r_Nc2}, $\rho(r)$ is plotted for $N=2$, $d=3,4,6,9$.
 They are all consistent with $c_{2,d}=c_{2,3}=2$.  
This should be related to the fact that $N=2, d>3$ theories are equivalent to a three-matrix model. 
We also found $c_{3,8}=17.2\pm0.5$ and $c_{3,9}=17.2\pm0.2$ to be compatible. However, that value does not agree with $(2\cdot 3-3)(8-1)=21$, presumably because of a possible $1/N$-correction explained above.

\newpage


\begin{thebibliography}{ccc}


\bibitem{Hawking:1976ra} 
  S.~W.~Hawking,  
  Phys.\ Rev.\ D {\bf 14}, 2460 (1976); 
Commun.\ Math.\ Phys. {\bf 87}, 395 (1982). 

\bibitem{Almheiri:2012rt} 
  A.~Almheiri, D.~Marolf, J.~Polchinski and J.~Sully,
  JHEP {\bf 1302}, 062 (2013)
  [arXiv:1207.3123 [hep-th]].

\bibitem{Braunstein:2009my}
Braunstein, Samuel L. and Pirandola, Stefano and Zyczkowski, Karol, 
PRL {\bf 110},101301, (2013) [arXiv:0907.1190[quant-ph]],
        
\bibitem{Maldacena:1997re}
  J.~M.~Maldacena, 
Adv.\ Theor.\ Math.\ Phys.\  {\bf 2}, 231 (1998).  

\bibitem{Itzhaki:1998dd}
  N.~Itzhaki, J.~M.~Maldacena, J.~Sonnenschein and S.~Yankielowicz,
Phys.\ Rev.\ D {\bf 58}, 046004 (1998).
  
  
\bibitem{Banks:1996vh}
T.~Banks, W.~Fischler, S.~H.~Shenker and L.~Susskind,
Phys.\ Rev.\ D {\bf 55}, 5112 (1997).     
  
\bibitem{deWit:1988ig} 
  B.~de Wit, J.~Hoppe and H.~Nicolai,
  Nucl.\ Phys.\ B {\bf 305}, 545 (1988).
 
 
\bibitem{Anagnostopoulos:2007fw} 
  K.~N.~Anagnostopoulos, M.~Hanada, J.~Nishimura and S.~Takeuchi,
  Phys.\ Rev.\ Lett.\  {\bf 100}, 021601 (2008). 

\bibitem{Catterall:2008yz}
  S.~Catterall and T.~Wiseman,
  Phys.\ Rev.\  D {\bf 78}, 041502 (2008). 

\bibitem{Kadoh:2015mka} 
  D.~Kadoh and S.~Kamata,
  arXiv:1503.08499 [hep-lat].

\bibitem{Filev:2015hia} 
  V.~G.~Filev and D.~O'Connor,
  arXiv:1506.01366 [hep-th].

\bibitem{Kabat:1999hp} 
  D.~N.~Kabat and G.~Lifschytz,
  Nucl.\ Phys.\ B {\bf 571}, 419 (2000)
  [hep-th/9910001].

\bibitem{Kabat:2000zv} 
  D.~N.~Kabat, G.~Lifschytz and D.~A.~Lowe,
  Int.\ J.\ Mod.\ Phys.\ A {\bf 16}, 856 (2001)
  [Phys.\ Rev.\ Lett.\  {\bf 86}, 1426 (2001)]
  [hep-th/0007051].
  
  
\bibitem{Hanada:2008ez} 
  M.~Hanada, Y.~Hyakutake, J.~Nishimura and S.~Takeuchi,
  Phys.\ Rev.\ Lett.\  {\bf 102}, 191602 (2009). 



\bibitem{Hanada:2013} 
M.~Hanada, Y.~Hyakutake, G.~Ishiki and J.~Nishimura,
Science 23 May 2014: Vol. 344 no. 6186 pp. 882-885. 


\bibitem{de Wit:1988ct} 
  B.~de Wit, M.~Luscher and H.~Nicolai,
  Nucl.\ Phys.\ B {\bf 320}, 135 (1989).

\bibitem{Banks:1997hz} 
  T.~Banks, W.~Fischler, I.~R.~Klebanov and L.~Susskind,
  Phys.\ Rev.\ Lett.\  {\bf 80}, 226 (1998)
  [hep-th/9709091].


\bibitem{Banks:1997tn} 
  T.~Banks, W.~Fischler, I.~R.~Klebanov and L.~Susskind,
  JHEP {\bf 9801}, 008 (1998)
  [hep-th/9711005].


\bibitem{Banks:1997cm} 
  T.~Banks, W.~Fischler and I.~R.~Klebanov,
  Phys.\ Lett.\ B {\bf 423}, 54 (1998)
  [hep-th/9712236].

 
\bibitem{Witten:1995im} 
  E.~Witten,
  Nucl.\ Phys.\ B {\bf 460}, 335 (1996)
  [hep-th/9510135].

\bibitem{Taylor:1999qk} 
  W.~Taylor,
  NATO Sci.\ Ser.\ C {\bf 556}, 91 (2000)
  [hep-th/0002016].

\bibitem{Taylor:2001vb} 
  W.~Taylor,
  Rev.\ Mod.\ Phys.\  {\bf 73}, 419 (2001)
  [hep-th/0101126].

\bibitem{Banks:1996my} 
  T.~Banks and N.~Seiberg,
  Nucl.\ Phys.\ B {\bf 497}, 41 (1997)
  doi:10.1016/S0550-3213(97)00278-2
  [hep-th/9702187].

\bibitem{Seiberg:1997ad} 
  N.~Seiberg,
  Phys.\ Rev.\ Lett.\  {\bf 79}, 3577 (1997)
  doi:10.1103/PhysRevLett.79.3577
  [hep-th/9710009].
  
\bibitem{Bigatti:1997jy} 
  D.~Bigatti and L.~Susskind,
  In *Cargese 1997, Strings, branes and dualities* 277-318
  [hep-th/9712072].
  
\bibitem{Banks:1999az} 
  T.~Banks,
  hep-th/9911068.

\bibitem{Joseph:2015xwa} 
  A.~Joseph,
  Int.\ J.\ Mod.\ Phys.\ A {\bf 30}, no. 27, 1530054 (2015)
  [arXiv:1509.01440 [hep-th]].

\bibitem{Hanada:2012eg} 
  M.~Hanada,
  PoS LATTICE {\bf 2012}, 019 (2012)
  [arXiv:1212.2814 [hep-lat]].

\bibitem{Catterall:2009it} 
  S.~Catterall, D.~B.~Kaplan and M.~Unsal,
  Phys.\ Rept.\  {\bf 484}, 71 (2009)
  [arXiv:0903.4881 [hep-lat]].


 
  
\bibitem{Kawahara:2007ib}
  N.~Kawahara, J.~Nishimura and S.~Takeuchi,
  JHEP {\bf 0712} (2007) 103
  [arXiv:0710.2188 [hep-th]].
  
  
\bibitem{Savvidy:1982jk} 
  G.~K.~Savvidy,
  Nucl.\ Phys.\ B {\bf 246}, 302 (1984).
  
  
\bibitem{Asplund:2011qj} 
  C.~Asplund, D.~Berenstein and D.~Trancanelli,
  Phys.\ Rev.\ Lett.\  {\bf 107}, 171602 (2011)
  [arXiv:1104.5469 [hep-th]].



\bibitem{Asplund:2012tg} 
  C.~T.~Asplund, D.~Berenstein and E.~Dzienkowski,
  Phys.\ Rev.\ D {\bf 87}, no. 8, 084044 (2013)
  [arXiv:1211.3425 [hep-th]].
 



\bibitem{Aoki:2015uha} 
  S.~Aoki, M.~Hanada and N.~Iizuka,
  JHEP {\bf 1507}, 029 (2015)
  [arXiv:1503.05562 [hep-th]].



\bibitem{Gur-Ari:2015rcq} 
  G.~Gur-Ari, M.~Hanada and S.~H.~Shenker,
  arXiv:1512.00019 [hep-th].




\bibitem{Sekino:2008he} 
  Y.~Sekino and L.~Susskind,
  JHEP {\bf 0810}, 065 (2008)
  [arXiv:0808.2096 [hep-th]].


\bibitem{Shenker:2013pqa} 
  S.~H.~Shenker and D.~Stanford,
  JHEP {\bf 1403}, 067 (2014)
  doi:10.1007/JHEP03(2014)067
  [arXiv:1306.0622 [hep-th]].


\bibitem{Shenker:2014cwa} 
  S.~H.~Shenker and D.~Stanford,
  JHEP {\bf 1505}, 132 (2015)
  doi:10.1007/JHEP05(2015)132
  [arXiv:1412.6087 [hep-th]].


  
\bibitem{Maldacena:2015waa} 
  J.~Maldacena, S.~H.~Shenker and D.~Stanford,
  arXiv:1503.01409 [hep-th].

\bibitem{Parikh:1999mf} 
  M.~K.~Parikh and F.~Wilczek,
  Phys.\ Rev.\ Lett.\  {\bf 85}, 5042 (2000)
  [hep-th/9907001].


\bibitem{Aoki:1998vn} 
  H.~Aoki, S.~Iso, H.~Kawai, Y.~Kitazawa and T.~Tada,
  Prog.\ Theor.\ Phys.\  {\bf 99}, 713 (1998)
  [hep-th/9802085].

\bibitem{Austing:2001bd} 
  P.~Austing and J.~F.~Wheater,
  JHEP {\bf 0102}, 028 (2001)
  [hep-th/0101071].

  
\end{thebibliography}
\end{document}